\documentclass[12pt,dvips]{article}
\usepackage{epsfig}

\newcommand{\condensate}{\ensuremath{\langle\bar{\psi}\psi\rangle}{}}
\newcommand\DWI{\ensuremath{D_\textrm{\scriptsize Wi}\,}{}}
\newcommand\DFP{\ensuremath{D_\textrm{\scriptsize Fp}\,}{}}
\newcommand\DOV{\ensuremath{D_\textrm{\scriptsize Ov}\,}{}}
\newcommand\unitmatrix{\ensuremath{\textrm{\boldmath{$\mathsf{1}$}}}}

\begin{document}

\thispagestyle{empty}

\date{\today}
\title{
\vspace{-5.0cm}
\begin{flushright}
{\normalsize UNIGRAZ-}\\
\vspace{-12pt}
{\normalsize UTP-16-05-01}\\
\end{flushright}
\vspace*{5cm}
Bound States for Overlap and Fixed Point Actions Close to the Chiral Limit}
\author
{\bf Stefan H\"ausler  and C. B. Lang \\  \\
Institut f\"ur Theoretische Physik,\\
Universit\"at Graz, A-8010 Graz, AUSTRIA}
\maketitle
\begin{abstract}
We study the overlap and the fixed point Dirac operators for massive fermions
in the two-flavor lattice Schwinger model. The masses of the triplet (pion) and
singlet (eta) bound states are determined down to small fermion masses and the
mass dependence is compared with various continuum model approximations. Near
the chiral limit, at very small fermion masses the fixed point operator has
stability problems, which in this study are dominated by finite size effects,
however.
\end{abstract}



\newpage
 
\section{Motivation and Introduction}

Lattice Dirac operators obeying the Ginsparg-Wilson condition \cite{GiWi82} in
its simplest form
\begin{equation}\label{gwc}
\gamma_5\,D+D\,\gamma_5=a\,D\,\gamma_5\,D\;,
\end{equation}
have the advantage to violate chiral symmetry only locally, with quasi
automatic ${\cal O}(a)$ improvement. Their eigenvalues lie on a circle in the
complex plane; zero eigenvalues correspond to chiral eigenstates and indicate
topological (instanton) modes in the gauge field. We known of at least two
explicit incarnations of such actions, the so-called overlap action
\cite{NaNe939495} and the  perfect action \cite{HaNi9498}. Both are technically
difficult and expensive to determine and to incorporate in a lattice simulation
with dynamical  fermions.

The Schwinger model (2D fermion-gauge theory with U(1) gauge group) generalized
to e.g. 2 flavors provides an attractive model to study such actions. It has
similar features like QCD with 2 quark flavors -- although due to a quite 
different underlying dynamics. It has only bosonic bound states, including a
massless triplet, and a non-vanishing fermion condensate. The overlap action
may be constructed from an arbitrary action, like the Wilson action. It
involves a computation of the sign function for hermitian matrices, which has
to be repeated for each gauge field and is computationally challenging. A
fixed point action (the perfect action determined at the fixed point of  the
continuum limit) on the other hand has a large number of  couplings and is
difficult to construct. For the Schwinger model such a fixed point action has been
determined in excellent approximation a few years ago \cite{LaPa98b}, at at
time, when the spectacular spectral properties of GW-fermion operators were not
yet widely discussed. Both, the overlap and the fixed point action then have been
studied in the Schwinger model for various aspects, mainly related to the Dirac
operator spectrum \cite{FaLaWo98} and to the topological content of the gauge
fields.

Introducing mass in the Schwinger model is a subtle problem 
\cite{CoJaSu75}-\cite{Sm97}
and the continuum approximations involve plausible but non-stringent 
ad-hoc assumptions \cite{Co76,Sm97}. Of particular interest is the analog of the PCAC-relation,
i.e. the dependence of the triplet bound state mass on the fermion mass
parameter. Recent studies for staggered action \cite{GuKaStWe98},
Wilson action \cite{GaHiLa99} and overlap action \cite{GiHoRe01}
have confirmed the exponent $2/3$. Here we try to obtain results closer
to the chiral limit for two GW-type actions. The motivation
is on one hand further confirmation of the functional dependence and 
a clearer determination of the correct coefficient. 

On the other hand we want to compare the relative merits of these chirally
improved actions closer to the chiral limit, which is hardly accessible with
e.g. Wilson-type actions. In fact, actions where (near the origin) the
eigenvalue spectra scatter have problems at small fermion masses in general.
Due to this fluctuation of the small real eigenvalues many configurations  will
mimic zero modes and necessitate special techniques to deal with them. The hope
is to improve this situation with overlap or perfect actions. We discuss our
conclusion based on this model study at the end.

\section{The massive 2-flavor Schwinger model}

We study the euclidean 2d gauge theory with gauge group U(1) and two
flavors of fermions with degenerate mass. For vanishing fermion masses, the
classical theory has a symmetry group $U(2)_L \times U(2)_R$ that is
broken down by the anomaly to $SU(2)_L \times SU(2)_R \times U(1)_V$. 
For non-vanishing fermion masses the chiral symmetry is broken at the
classical level and a $U(2)_V$ symmetry remains. 

Let us briefly discuss some known analytical results for the continuum model.
The generalized Sine-Gordon model bosonizes the
2-flavor Schwinger model\cite{Co76}. Its Lagrangian reads
\begin{equation}
\label{gensg}
\frac{1}{2}\left(\partial \varphi^{(0)}\right)^2 +
\frac{1}{2}\left(\partial \varphi^{(3)}\right)^2 +
\frac{1}{2}\, \frac{2\,g^2}{\pi} \left(\varphi^{(0)}\right)^2 
- \frac{m \,c}{\pi}\, \cos\left(\sqrt{2 \pi} \,\varphi^{(0)}\right) 
\cos\left(\sqrt{2 \pi} \,\varphi^{(3)}\right) \; .
\end{equation}
The constant $c$ is not determined by the bosonization but is related  (see
e.g.~\cite{GaSe94,Ga96}) to the masses $\mu^{(0)}$ and $\mu^{(3)}$ used for
normal ordering of the fields  $\varphi^{(0)}$ and $\varphi^{(3)}$ by  $c =
(\mu^{(0)} \mu^{(3)})^{1/2} e^\gamma /2$, where $\gamma$ ($= 0.577216\ldots$) 
denotes Euler's
constant. For the case of two flavors two of
the currents -- which we measure -- (compare (\ref{currents})) are bosonized with the
following prescription 
\begin{equation}
\overline{\psi}(x) \; \tau^a \otimes \gamma_\mu \; \psi(x)
\; = \; \frac{1}{\sqrt{\pi}}\, \varepsilon_{\mu \nu}\, \partial_\nu \;
\varphi^{(a)}(x) \quad , \qquad a = 0,3 \; ,
\label{bosonization}
\end{equation}
where the generators $\tau^a$ of rotations in flavor space are given by 
the Pauli matrices $\tau^a = \sigma_a, \; a = 1,2,3$ and $\tau^0 \equiv$
1\hspace{-1.4mm}1.
For the other two members $j_\mu^{(1)}, j_\mu^{(2)}$ of the iso-triplet no
explicit abelian bosonization is known; 
however due to invariance
under flavor rotations, their masses are the same as for the triplet current
($a=3$, which we called pions due to the analog to QCD). 
For vanishing quark mass $m$  in (\ref{gensg}) the
two flavor Schwinger model is bosonized by two free fields. One of them, the
pion field $\varphi^{(3)}$, is massless ($M_\pi = 0$) and the eta field 
$\varphi^{(0)}$ obtains the Schwinger mass $M_\eta = g \sqrt{2/\pi}$
(due to the anomaly).

For non-vanishing quark mass, also the bosonized model (\ref{gensg}) can no
longer be solved in closed form. A semi-classical
analysis  (see e.g.~\cite{Ga96}) should provide a good approximation when all
involved masses are large. The
squares of the masses are given by the second derivatives of the 
interaction part
$V[\varphi^{(0)},\varphi^{(3)}]$ of (\ref{gensg}) at the minimum.  After normal ordering the 
fields $\varphi^{(a)}$ with respect to their own masses (setting 
$\mu^{(0)} = M_\eta$ and $\mu^{(3)} = M_\pi$), the semi-classical analysis for
the iso-triplet and iso-singlet gives
\begin{eqnarray}
\label{semitrip}
\frac{M_\pi}{g} \; &=&  \; e^{2\gamma/3}\; \frac{2^{5/6}}{\pi^{1/6}} \;  
\left(\frac{m}{g}\right)^{2/3}\\
\label{semising}
\frac{M_\eta}{g} \; &=& \; \sqrt{\frac{2}{\pi} \; + \; 
\left(\frac{M_\pi}{g}\right)^2}\; .
\end{eqnarray}
The first relation plays a role similar to the PCAC relation in QCD
(where the pion mass squared is proportional to the quark mass).

In an attempt to go beyond semi-classical analysis one can approximate  the
generalized Sine-Gordon model (\ref{gensg}) by a solvable model.
In the  limit of large coupling \cite{Co76}
$g$ and small mass $m$ the iso-singlet field $\varphi^{(0)}$ becomes
static and the model is reduced to a standard  Sine-Gordon model for the
triplet field $\varphi^{(3)}$, with the Lagrangian
\begin{equation}
\frac{1}{2}\left(\partial \varphi^{(3)}\right)^2 
\; - \; 2C \cos\left(\sqrt{2 \pi} \varphi^{(3)}\right) \; .
\label{standardsg}
\end{equation}
This reduced bosonized theory has been studied \cite{Co76,Gr91} using the WKB
approximation \cite{DaHaNe75} and recently again by Smilga \cite{Sm97}. His
analysis is now based on the newly derived  analytic solution  \cite{Za95} of
the standard Sine-Gordon model and he finds 
\begin{equation}
\label{smilgatrip}
\frac{M_\pi}{g} \; = \; 2^{5/6}\, e^{\gamma/3} \,
\left( \frac{\Gamma(\frac{3}{4})}{\Gamma(\frac{1}{4})} \right)^{2/3}
\frac{\Gamma(\frac{1}{6})}{\Gamma(\frac{2}{3})} \; 
\left(\frac{m}{g}\right)^{2/3}\; .
\end{equation} 
The numerical coefficient of the $(m/g)^{2/3}$ term is 2.008, somewhat smaller
than the value 2.1633 in (\ref{semitrip}).
The truncation of the original model (\ref{gensg}) to the 
standard form (\ref{standardsg}) does not allow for a result for the 
singlet mass $M_\eta$. For this state only the semi-classical  formula
(\ref{semising}) is available. It his however interesting to use the exact
result (\ref{smilgatrip}) of the  truncated model as an input in
(\ref{semising}), and below we compare also this formula to the numerical
data.

\section{Lattice Dirac Operators}

The Schwinger model is a super-renormalizable theory, i.e.~the  bare
coupling does not get renormalized and equals the physical gauge coupling $g$. On
the lattice for the standard Wilson plaquette gauge action we use the usual
coupling $\beta$ in front of the gauge field action. In the naive  continuum
limit this is related to the continuum coupling via   $\beta=1/g^2$ (we set
the lattice spacing $a = 1$). We define the coupling $g$ which we use for
comparison with the analytical  results in the continuum (see above) through
this relation. 

We first generate the massless fermion actions and introduce the  fermion
mass parameter subsequently.

\subsubsection*{Overlap Dirac operator}

For the overlap action \cite{Ne98} one may start with the Wilson 
Dirac operator
\begin{equation}\label{DWI}
\DWI(x,y)=(m_w+2)\,\delta_{xy} - \frac{1}{2} \sum_{\mu=\hat{1},\hat{2}}
 \left[(1+\sigma_\mu) \,U_{xy}\,\delta_{x,y-\mu}
+ (1-\sigma_\mu)\, U_{yx}^\dagger\,\delta_{x,y+\mu}\right]\;,
\end{equation}
(with the Pauli matrices $\sigma_\mu$)
at some value of $m_w \in (-1,0)$ and then construct 
\begin{equation}\label{DOV}
\DOV = \frac{1}{2}\left[ \unitmatrix  + \gamma_5\, \epsilon(\gamma_5\,\DWI) \;\right].
\end{equation}
Some words about the choice of $m_w$: according to \cite{Ne98} it is
arbitrary, in the sense that any (strictly negative) value in the interval
$(-1,0)$ reproduces the correct continuum theory, but it may be optimized with
regard to its scale dependence by looking for example at the behavior of the
(projected) spectrum. In many practical determination of the sign function
configurations with small eigenvalues of $\gamma_5\,\DWI$  are cumbersome;
this situation may be improved by a suitable choice of $m_w$. Comparing
expectation values of operators like $\condensate$ for different $m_w$ one
has to take care of the proper normalization \cite{KiNaNe98}. 
Comparing with free lattice fermions one finds a (trivial)
factor of $\sqrt{2\,m_w}$  renormalizing the field operator, i.e. 
$\condensate=Z_\psi^{-1}\condensate_\textrm{\scriptsize Ov}$ with
$Z_\psi=|2\,m_w|$ in our convention. We choose $m_w=-1$.

The operative definition of  the generalized sign function
$\epsilon(\gamma_5\,\DWI)$ entering the above equation is through its
eigenvalues,
\begin{equation}\label{defeps}
\epsilon(\gamma_5\,\DWI)= 
U \, \textrm{Sign}(\Lambda)\, U^\dagger \quad \textrm{with}\; 
\gamma_5\,\DWI = U \, \Lambda \,U^\dagger\;.
\end{equation}
Here $\textrm{Sign}(\Lambda)$ denotes the diagonal matrix containing  the
signs of the eigenvalues obtained through the unitary
transformation $U$ of the hermitian  matrix $\gamma_5\,\DWI$. There are
various ways to numerically find \DOV without passing through  the
diagonalization problem (which is  prohibitively expensive for $D=4$) 
\cite{Ne98e,EdHeNa98c,HeJaLe9900} (for $D=2$ see also \cite{Ch98} and
recently, in the Schwinger model \cite{GiHoRe99,GiHoRe01}). In our simple
context computer time is no real obstacle and therefore we use the direct
definition (\ref{defeps}), explicitly performing the diagonalization. For
comparison we also applied an iterative  technique implementing Newton's
method \cite{Ne98e,Ch98}. For many gauge configurations the resulting
operators agreed to the requested accuracy (8 digits). For larger $\beta\ge 4$
however, the number of configurations with convergence problems for  the
iterative method due to small eigenvalues increased. The results presented here
are all based on the exact  diagonalization.

\subsubsection*{Fixed point Dirac operator}

In Ref.\cite{LaPa98b} the fixed point Dirac operator was parameterized as
\begin{equation}\label{DFP}
\DFP(x,y) =
\frac{1}{2}\,\sum_{i=0}^3\sum_{x\, , f}\, \rho_i(f)\, 
\sigma_i\, U(x,f)\;,\quad\textrm{with} \; y\equiv x+\delta f\;.
\end{equation}
Here $f$ denotes a closed loop through $x$ or a path from the lattice site
$x$ to $y=x+\delta f$ (distance vector $\delta f$) and $U(x,f)$ is the
parallel transporter along this path. The $\sigma_i$-matrices denote the
Pauli matrices for $i= 1,2,3$ and the unit matrix for $i=0$.  Note the
factor $1/2$ in (\ref{DFP}) in order to obtain the same normalization as the overlap
Dirac operator. The action obeys the usual symmetries as discussed in
\cite{LaPa98b}; altogether it has 429 terms per site. The action was
originally determined at large $\beta$ for gauge fields distributed
according to the non-compact formulation with the Gaussian measure. 
Excellent  scaling properties, rotational invariance and continuum-like
dispersion relations were observed at various smaller values of the gauge
coupling $\beta$.

In Ref.\cite{FaLaWo98} this action was studied both, for compact and the original
non-compact gauge field distributions.  In the compact case the action is not
expected to exactly reproduce the fixed point of the corresponding BST, but
nevertheless it is still a solution of the GWC (\ref{gwc}); violations
are instead introduced by the parameterization procedure, which cuts off
the less local couplings. It was demonstrated, that  the eigenvalue spectrum is
close to circular shape, improving in this respect towards larger $\beta$ (see
\cite{FaHiLaWo98} for a more detailed discussion).

Here we study the action only for the compact gauge field distributions in
order to allow for a direct comparison with the overlap Dirac operator.

\subsubsection*{The massive case}

The massive overlap Dirac operator may be related to the massless one by
\cite{EdHeNa99c}
\begin{equation}\label{massive DOV}
\DOV(\mu)=(1-\mu)\left[ \DOV(0) + \frac{\mu}{1 - \mu}\right]\;,
\end{equation}
where the mass parameter $-1 < \mu <1$ is related to the fermion mass by 
\begin{equation}
m=\mu\,Z_m^{-1}(1+\mathcal{O}(a^2)),
\end{equation}
with $Z_\psi Z_m=1$, where $Z_m$ and $Z_\psi$ are the mass and the
wavefunction renormalization constants, respectively. Since with our choice
of parameterization the massless fixed point Dirac operator (\ref{DFP}) has
the same overall renormalization as the massless overlap Dirac operator
(\ref{DOV}) we also use  (\ref{massive DOV}) to generate the massive fixed
point Dirac operator.

\section{Simulation Details}

Uncorrelated gauge configurations for lattice size $16^2$ and $24^2$ have
been generated in the quenched setup.  However, we are including the
fermionic determinant in the observables:  all the results presented here
are obtained  with the correct determinant (squared, for two flavors)
weight. From earlier experience \cite{LaPa98b,FaLaWo98} we learned that this
is justifiable for this model and the presented statistics.  We perform our
investigation on two sets of 5000-10000 configurations at $\beta=4$ and 6.
These are the same gauge configurations as used in Ref.\cite{GaHiLa99} for the
determination of bound state masses for the Wilson Dirac operator and thus
we may compare the relative efficiency in the results.

The configurations have been well separated by  three times the
autocorrelation length for the so-called ``geometric definition'' of  the
topological charge. For each configuration we construct \DOV and \DFP as
discussed.  For each lattice Dirac matrix we then compute the inverse (the
quark propagator)  and the determinant.

The masses for the pions and the eta-particle were determined from the 
decay of two point functions of the following {\em vector currents}
\begin{eqnarray}
J_x^{a \mu} & = & \overline{\psi}_x \; \tau^a \otimes \gamma_\mu \; \psi_x
\; \; \; \; \; \mbox{for the $\pi$'s} \; , 
\nonumber \\
J_x^{0 \mu} & = & \overline{\psi}_x \; \mbox{1\hspace{-1.4mm}1} 
\otimes \gamma_\mu \; \psi_x \; \; \; \; \; \; \; \mbox{for the} \;
\eta \; .
\label{currents}
\end{eqnarray} 
We emphasize that 2-point functions of {\em scalars} or  {\em
pseudo-scalars} are not  particularly well suited for the determination of
the meson masses. These operators are bosonized (compare the discussion
above) by cosines and  sines of the fundamental fields
\cite{BeSwRoSch79,GaSe94} and their 2-point  functions strongly mix
contributions from both the triplet and singlet states.  For the two point
functions of the momentum-zero projected {\em vector currents} one has to  take care
about the proper choice of $\mu$. As could be verified by gauge field
integration there are no contributions to the propagation of $J_x^{a
2}(p=0)$ in the direction $\mu=2$.

\begin{figure}[htpb]
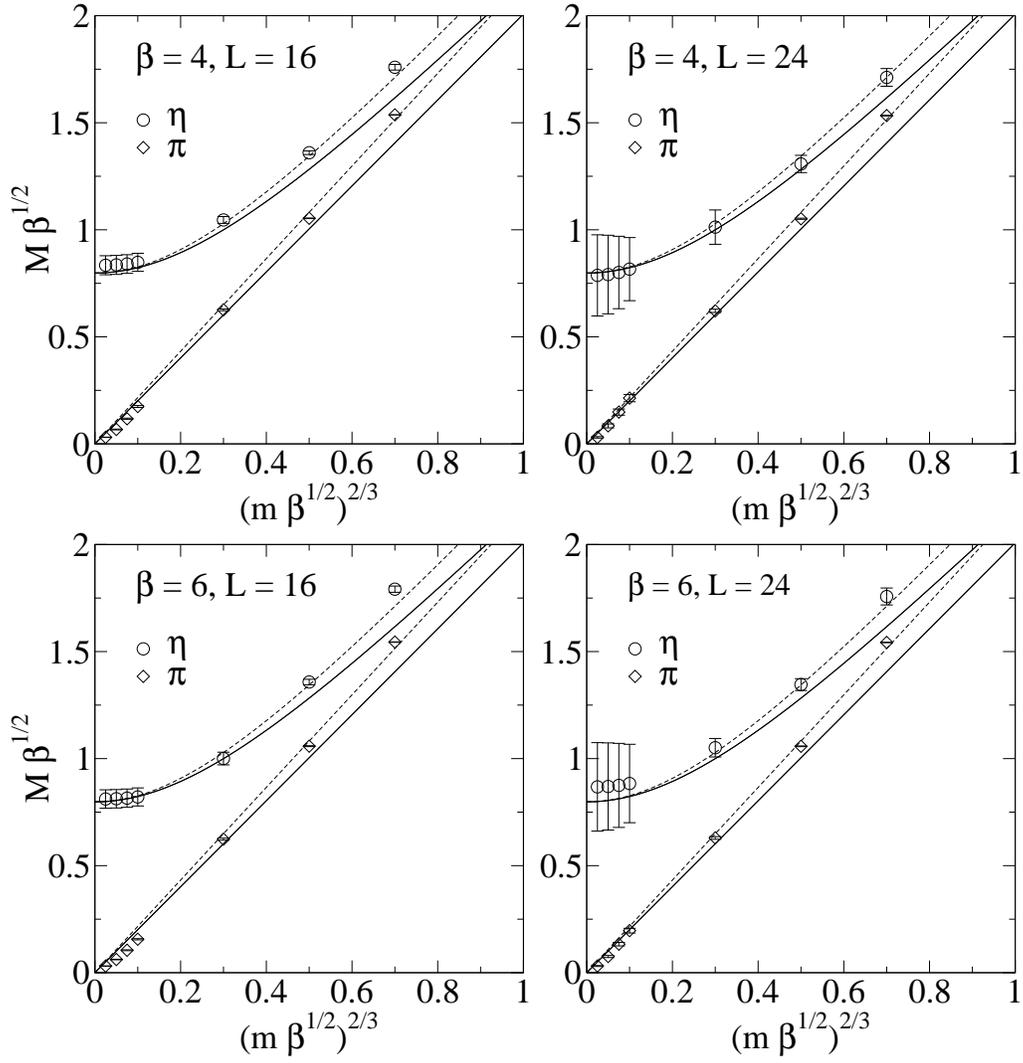

\epsfig{file=Nb_L16b04.eps,height=7.0cm,clip=}
\epsfig{file=Nb_L24b04.eps,height=7.0cm,clip=}\\
\epsfig{file=Nb_L16b06.eps,height=7.0cm,clip=}
\epsfig{file=Nb_L24b06.eps,height=7.0cm,clip=}\\
\caption{Results for overlap Dirac operator $\beta=4$, 6, $L=16$, 24.
Symbols: Monte Carlo data. Full line: Smilga's formula 
(\protect{\ref{smilgatrip}}). Dashed line:
the semi-classical approximation
(\protect{\ref{semitrip}}) and (\protect{\ref{semising}}).
\label{overlapplot}}
\end{figure}

\begin{figure}[htpb]
\epsfig{file=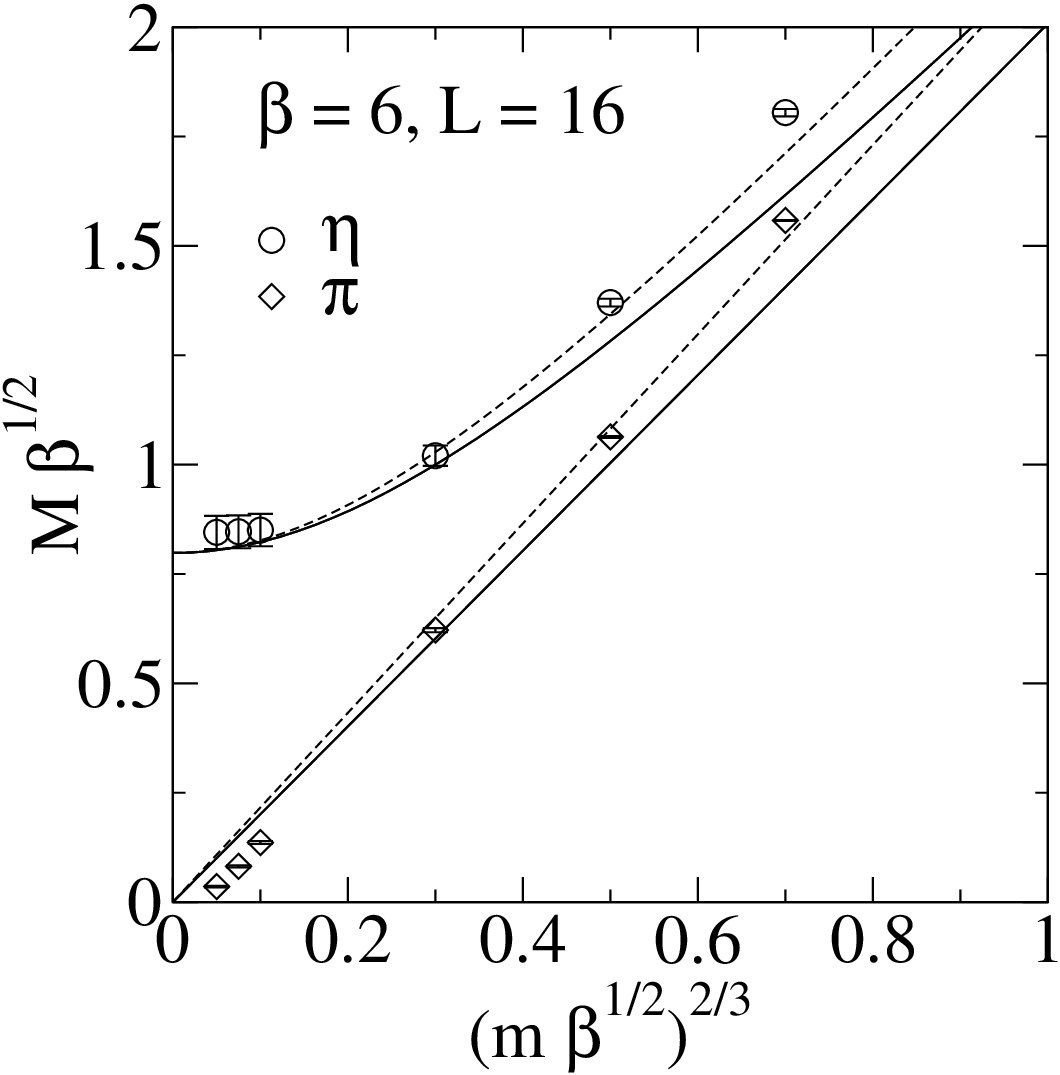,height=7.0cm,clip=}
\epsfig{file=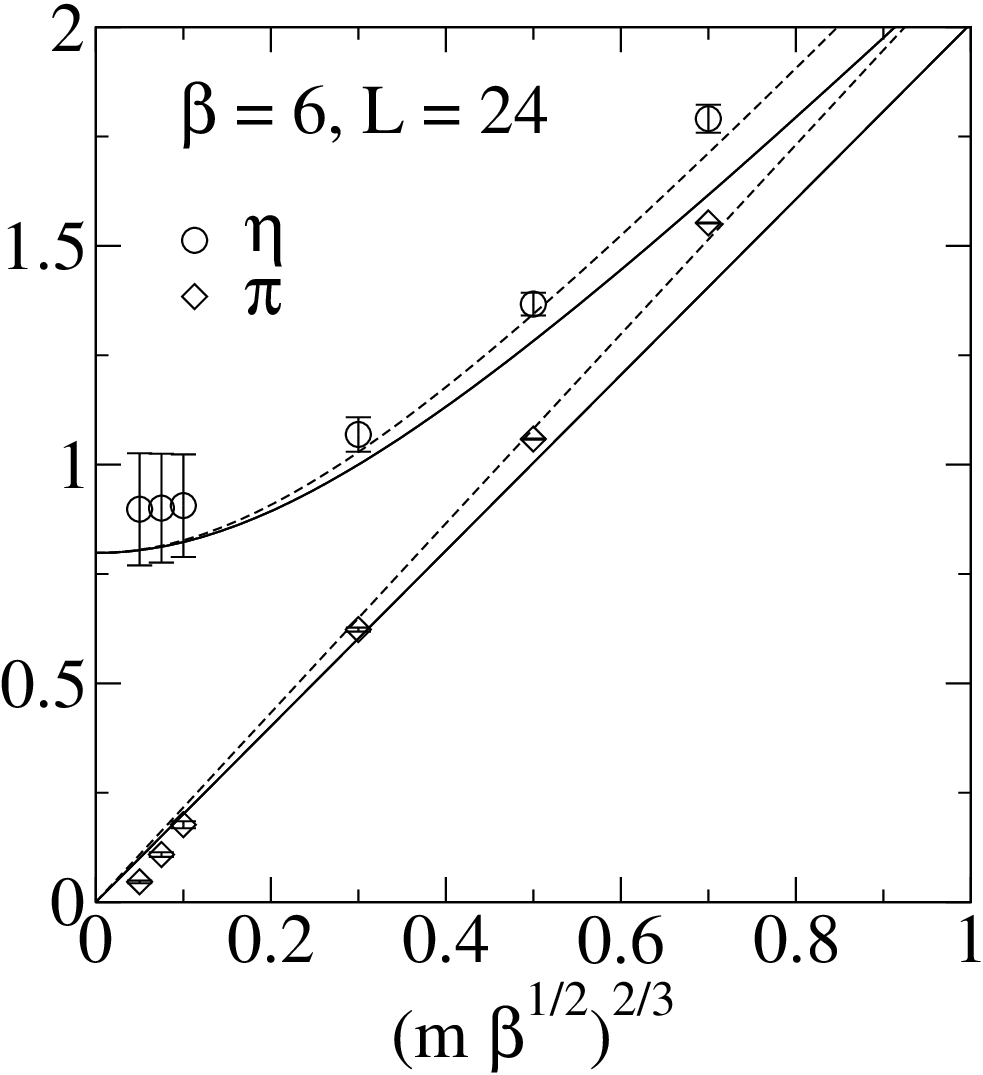,height=7.0cm,clip=}\\
\caption{Results for the fixed point Dirac operator at $\beta=6$, $L=16$, 24. 
Symbols: Monte Carlo data. Full line: Smilga's formula 
(\protect{\ref{smilgatrip}}). Dashed line:
the semi-classical approximation
(\protect{\ref{semitrip}}) and (\protect{\ref{semising}}).
\label{fixedpointplot}}
\end{figure}

\section{Results}

Let us now discuss our numerical results for the mass spectrum.   We present
the data for the overlap Dirac operator  for $\beta = 4$ and 6 and $L=16$  and
24 in Fig.~\ref{overlapplot}. The results for the fixed point operator at
$\beta = 6$ and $L=16,24$ are shown in Fig.~\ref{fixedpointplot}. In each plot
we combine the values of the singlet mass and the triplet mass and  compare the
numerical results to Smilga's formula (\ref{smilgatrip}) (full lines) and the
semi-classical results (\ref{semitrip}) and (\ref{semising})(dotted lines).

We observe that the data are well described by the semi-classical result
(\ref{semitrip}) (dotted line). In particular at quark masses above
$(m\sqrt{\beta})^{2/3} \ge 0.3$ the semi-classical ap\-proximation is
fulfilled best. In this region our results for the triplet mass coincide 
with the results of \cite{GiHoRe01}. For small
quark masses we observe a strong and seemingly systematic deviation 
especially for small lattice size. We attribute this to finite size
effects. In particular, the boson propagator does not really become
asymptotic at such large correlation lengths.

For the fixed point operator we only get reasonably results for small $m$ at
large gauge couplings $\beta \ge 4$, when the spectrum is close to circular.
For smaller $\beta$ the scattering of the small eigenvalues is too large and
of the order of the bare fermion mass parameter, causing disturbing
fluctuations. The dispersion $\sigma$ of the eigenvalues around the circle
(which decreases with larger $\beta$\cite{FaHiLaWo98})
leads to accidental poles  in the propagators even at non-vanishing quark
mass.

For the eta mass $M_\eta$ we find that the data approach the correct 
$m=0$ value as the quark mass vanishes. The semi-classical formula
(\ref{semising}) provides a reasonable  description of the data. For small
$m$ the quantum fluctuations become larger  and thus the numerical data
deviate from the semi-classical curve. The cases with comparatively large
error bars always indicate lack of sufficient statistics for the 
propagator. The behavior at small $m$ can be fitted by a power law as has
been done  in Ref.\cite{GuKaStWe98} for the results from the model with
staggered fermions.
 
Comparing the results with those obtained in Ref.\cite{GaHiLa99} for the Wilson 
Dirac operator we find generally smoother behavior, in particular at smaller
fermion mass parameters. This was to be expected due to the improved
chirality  properties. On the other hand, for a full exploitation of this
feature larger lattices with better finite size control are needed.

\section{Conclusion}

We find that our results for the singlet mass and the triplet mass are well
described by the semi-classical formulas (\ref{semitrip}), (\ref{semising}). 
For the triplet
mass $M_\pi$ we observe a clear deviation from the semi-classical curve at
small quark masses which we attribute to finite volume effects. For the 
$\eta$-mass the data are in good agreement with the  semi-classical
approximantion with quantum fluctuations becoming less important
as the quarks are made heavier. 

As for the studied model, although the results for $\eta$ and $\pi$ are in
good agreement with the the analytic expectations the values of the pion
mass below  $M_\pi \sqrt{\beta} < 0.3$ are clearly plagued by the limited
lattice size. However,  our emphasis was on the study of the effects due to the
chiral properties of the Dirac operators. For better results for the 
$\pi$-mass at very small masses much larger lattice and higher statistics appear
to be necessary to allow a control of the finite size effects.

For small fermion masses it seems to be important, that the  eigenvalues of
the Dirac operator do not scatter too much away from the optimal unit
circle. With this regard the overlap action does the better job. Even the
qualitatively quite good fixed point action exhibits a weakness for small
$\beta$ due to this effect. Spurious singularities or almost-singularities in
the propagator necessitate very large samples in order to obtain reliable
mean expectation values and thus propagators. The width of the distribution
of eigenvalues around zero therefore essentially limits the practically
applicable values of the bare fermion mass parameter.

{\bf Acknowledgment:} We want to thank C. Gattringer and I. Hip for many useful
discussions.

\end{document}